\documentclass[preprint]{elsarticle}

\usepackage{amssymb}

\usepackage[utf8]{inputenc} 
\usepackage[T1]{fontenc}    
\usepackage{amsmath}
\usepackage{url}            
\usepackage{booktabs}       
\usepackage{amsfonts}       
\usepackage{nicefrac}       
\usepackage{microtype}      
\usepackage{graphicx}
\usepackage{multirow}
\usepackage{wrapfig}
\usepackage{lscape}
\usepackage{xcolor}
\usepackage{subfig}
\usepackage{lscape}
\usepackage{rotating}
\usepackage{epstopdf}
\usepackage{pdfpages}
\usepackage{framed} 
\usepackage{multicol} 
\usepackage{nomencl} 
\usepackage{float}
\usepackage{wrapfig}
\usepackage[colorlinks=true]{hyperref}
\makenomenclature
\renewcommand*\nompreamble{\begin{multicols}{2}}
\renewcommand*\nompostamble{\end{multicols}}

\journal{Journal of Hydrology: Regional Studies}

\begin{document}
\hypersetup{
    urlcolor=magenta,
    citecolor=blue,
    linkcolor=red
    }
\begin{frontmatter}

\author{Sakshi Dhankhar\corref{cor1}}
\ead{ss97@tu-clausthal.de}
\author{Stefan Wittek}
\ead{stefan.wittek@tu-clausthal.de}
\author{Hamidreza Eivazi}
\ead{hamidreza.eivazi.kourabbaslou@tu-clausthal.de}
\author{Andreas Rausch}
\ead{andreas.rausch@tu-clausthal.de}

\cortext[cor1]{Corresponding author at: Clausthal University of Technology, Institute for Software and Systems Engineering, Arnold-Sommerfeld Str. 1, 38678, Clausthal-Zellerfeld, Germany.}

\affiliation{organization={Clausthal University of Technology, Institute for Software and Systems Engineering},
            addressline={Arnold-Sommerfeld Str. 1}, 
            city={Clausthal-Zellerfeld},
            postcode={38678}, 
            country={Germany}}

\title{A Spatiotemporal Radar-Based Precipitation Model for Water Level Prediction and Flood Forecasting}

\begin{abstract}
\textit{Study Region:} Goslar and Göttingen, Lower Saxony, Germany.\\
\textit{Study Focus:} In July 2017, the cities of Goslar and Göttingen experienced severe flood events characterized by short warning time of only 20 minutes, resulting in extensive regional flooding and significant damage. This highlights the critical need for a more reliable and timely flood forecasting system. This paper presents a comprehensive study on the impact of radar-based precipitation data on forecasting river water levels in Goslar. Additionally, the study examines how precipitation influences water level forecasts in Göttingen. The analysis integrates radar-derived spatiotemporal precipitation patterns with hydrological sensor data obtained from ground stations to evaluate the effectiveness of this approach in improving flood prediction capabilities. \\
\textit{New Hydrological Insights for the Region:} A key innovation in this paper is the use of residual-based modeling to address the non-linearity between precipitation images and water levels, leading to a Spatiotemporal Radar-based Precipitation Model with residuals (STRPMr). Unlike traditional hydrological models, our approach does not rely on upstream data, making it independent of additional hydrological inputs. This independence enhances its adaptability and allows for broader applicability in other regions with RADOLAN precipitation. The deep learning architecture integrates (2+1)D convolutional neural networks for spatial and temporal feature extraction with LSTM for timeseries forecasting. The results demonstrate the potential of the STRPMr for capturing extreme events and more accurate flood forecasting.
\end{abstract}


\end{frontmatter}

\section{Introduction}
Floods rank among the most prevalent natural disasters, leading to considerable loss of life, widespread property destruction, and substantial economic disruption on a global scale. The development of precise early warning systems is crucial for effective disaster management and risk reduction, especially as climate change intensifies the frequency and severity of extreme weather events. According to the United Nations (UN), approximately one in three individuals do not have access to adequate multi-hazard early warning systems \cite{nations_early_nodate}. Currently, the pinnacle of real-time, global-scale hydrological prediction is represented by the Global Flood Awareness System (GloFAS) \cite{GloFAS,alfieri_glofas_2013,harrigan_daily_2023}, which forms the core of the flood forecasting operations within the Copernicus Emergency Management Service (CEMS). This system is established under the guidance of the European Commission's Joint Research Centre (JRC) and is operated by the European Centre for Medium-Range Weather Forecasts (ECMWF) as the central computational hub for hydrological forecasting tasks.

Traditional flood prediction models often depend on hydrological and meteorological data collected from rain gauges, stream gauges, and other ground-based sensors. The most common models are the Soil Conservation Service (SCS) Curve Number Method \cite{michel2005soil}, Hydrologic Engineering Center's Hydrologic Modeling System (HEC-HMS) \cite{scharffenberg2008hydrologic}, and the Soil and Water Assessment Tool (SWAT) \cite{gassman2007soil}. They are typically driven by inputs from ground-based weather stations and historical hydrological data and rely on deterministic or statistical methods to simulate water flow and predict river discharge. While these hydrological models are well-established models, regions that are most susceptible to flooding often still lack dependable forecasting and early warning infrastructure. Moreover, a common challenge is the limited availability of upstream hydrological data, which can hinder accurate predictions.

The increasing availability of large datasets has led to the development of data-driven models that leverage machine learning (ML) algorithms to forecast water levels. Techniques such as artificial neural networks (ANNs), support vector machines (SVMs), and long short-term memory (LSTM) networks have been widely explored in recent years. Recent advancements in artificial intelligence (AI), coupled with the increasing availability of open datasets, have facilitated the creation of data-driven models that considerably enhance the accuracy and lead time of short-term forecasts for extreme riverine events \cite{nearing_global_2024}. AI-based forecasting models have on average improved the reliability of global nowcasts and elevated the forecasting proficiency in Africa to levels comparable to those found in Europe \cite{nearing_global_2024}.

Remote sensing technologies, such as satellite and radar systems, have transformed flood prediction by providing more precise precipitation measurements. Several notable studies have demonstrated the potential of radar data in flood forecasting and water level prediction.~\citet{WANG2015408} showed that radar-based models could achieve better spatial representation of rainfall, leading to improved runoff simulations. A study by \citet{rs13163251} implemented radar data into the Weather Research and Forecasting (WRF) hydrological modeling system (WRF-Hydro), resulting in improved forecast skill during extreme rainfall events. Similarly, \citet{w16040607} employed deep-learning (DL) models on radar-derived precipitation data for real-time flood forecasting, demonstrating that DL techniques could further enhance prediction accuracy by capturing non-linear relationships within the data. Despite these advances, many existing spatial-temporal models focus on either spatial or temporal dimensions, rather than both in a unified framework.

Recent advancements in AI-based weather forecasting models, e.g. FourCastNet \cite{pathak_fourcastnet_2022}, GraphCast \cite{lam_graphcast_2023}, and Pangu-Weather \cite{bi_pangu-weather_2022}, showcase the effectiveness of machine learning architectures in handling complex dynamics of the atmosphere. By leveraging global datasets and sophisticated neural network architectures, these models can outperform traditional numerical weather prediction approaches in both accuracy and computational efficiency. Notably, these advancements include the development of foundation models such as ClimaX \cite{nguyen_climax_2023} and Aurora \cite{bodnar_aurora_2024}, which aim to provide generalizable solutions that can be adapted to a variety of atmospheric prediction tasks. These models illustrate how AI can be harnessed to improve the prediction of extreme weather events, including those linked to flooding. For instance, ClimaX leverages transformer architectures to address diverse climate and weather tasks, suggesting parallel applications for flood forecasting where understanding atmospheric precursors is critical. Furthermore, the physics-informed approaches proposed by Wang et al. \cite{wang_beyond_2024} and the integration of machine learning with general circulation models \cite{kochkov_neural_2024} highlight the potential to enhance flood forecasting systems by capturing both large-scale climatic influences and small-scale hydrological responses. The success and operationalization of these models can provide more timely and accurate flood warnings, especially in data-scarce regions, by addressing the challenges posed by chaotic weather systems \cite{mukkavilli_ai_2023}. This convergence of AI and environmental science holds the promise to transform our resilience to flooding and other climate-related hazards.

The authors in \cite{bratzel2023flood} introduced a flood prediction benchmark designed to assess machine-learning models in forecasting extreme hydrological events in Goslar. Their study emphasizes the challenges associated with predicting rare and severe flooding incidents, leveraging a residual-LSTM model trained on hydrological sensor data. By focusing on water level forecasting at a specific sensor location, their benchmark establishes a standard evaluation framework for comparing different predictive approaches. This work serves as a crucial baseline against which novel methodologies such as our proposed models STRPM and STRPM\textsubscript r can be quantitatively evaluated.

The AI-based models discussed so far are global in scope and provide valuable insights into flood forecasting on a broad scale. However, there is a crucial need to develop regional models that can deliver more precise forecasts tailored to specific areas, especially those that are highly susceptible to flooding and have limited availability of upstream data. By focusing on localized historical data and regional specificities, these models can achieve higher accuracy in predictions compared to broader, global frameworks. This highlights the need for regional models to complement global systems, ensuring more precise and timely flood warnings.
 
In this work, we utilize the precipitation data provided by the Deutscher Wetterdienst (DWD), which is collected through their specialized precipitation measurement network as well as equivalent partner networks. DWD's current radar system has 17 weather stations across Germany, which provides information about precipitation values. The radar precipitation values cannot be directly calculated from the ground, but they are rather calibrated using precipitation gauges (ombrometers) \cite{dwd}. The DWD precipitation data enables the development of highly accurate and finely-tuned regional models for flood forecasting. By utilizing this high-resolution and extensive dataset, we are able to identify localized weather patterns and variations, offering deeper insights into extreme weather events.

This paper introduces the Spatial-Temporal Radar-based Precipitation Model (STRPM), a novel approach designed to integrate radar-derived precipitation data with river water levels measured in Goslar. By capturing the dynamic nature of precipitation events, STRPM aims to improve the lead time and reliability of flood forecasts. This model addresses the limitations of limited upstream data by incorporating high-frequency radar observations, which offer detailed insights into the spatial distribution and intensity of rainfall. The primary objectives of this research are to develop and validate the STRPM, assess its performance in various scenarios, and demonstrate its potential for operational flood forecasting. The model's performance is systematically evaluated using the original Goslar Benchmark \cite{bratzel2023flood} together with additional experiments to assess its accuracy. To validate the model's robustness, additional testing was conducted on data from Göttingen. This validation demonstrated the model's flexibility, indicating its potential for future implementation in regions with varying topologies through the application of transfer-learning techniques.

In section \ref{methodology} we discuss the methodology and data sources, in section \ref{model arc}, the architecture used in the development of STRPM. We also present the results of our case studies in section \ref{results}, showcasing the model’s capability to predict water levels and forecast floods with high precision. Finally, we explore the implications of our findings for future research and practical applications in flood risk management in section \ref{conclusion}.

\section{Methodology} \label{methodology}
\subsection{Study Area}
A small historic town \textbf{Goslar} is located in the northwestern slopes of the Harz mountainous regions of Germany, which has two small rivers Gose and Abzucht converging and flowing into the city. Abzucht with a total catchment area of 31.6 km\textsuperscript{2}, stretching over 12.1 km in length descends a total of 188 metres and represents an average incline of 15.5\%. Gose is a left tributary of Abzucht, stretching over 7.1 km with a drainage basin of 10 km\textsuperscript{2}.  The stream flows steadily southwards and flows into the Abzucht at the southern outskirts of Goslar \cite{niedersachsen}. Goslar is located 15 km northwest of Broken, which is recorded as the highest peak in Harz. This UNESCO world heritage site was flooded by these two rivers on July 26\textsuperscript{th}, 2017, and around 120 people were evacuated \cite{floodlist}. 

\begin{figure}[h]
    \centering
    \includegraphics[width=1\textwidth]{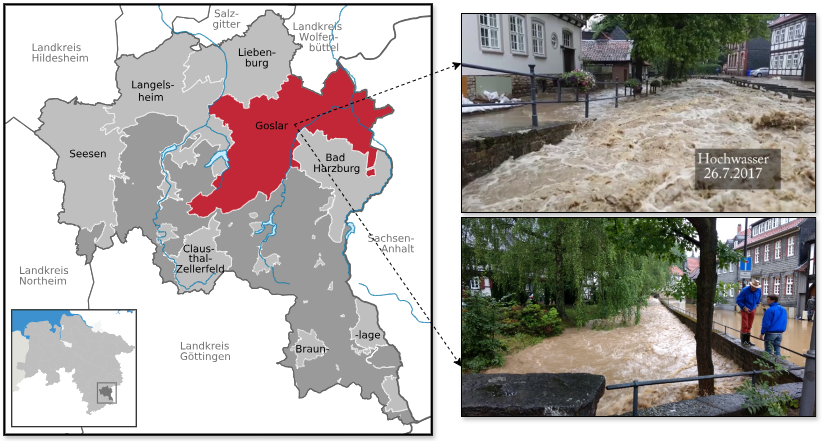}
    \caption{Location overview of Goslar in Lower Saxony, Germany \cite{GoslarSVG}.}
    \label{fig:map}
\end{figure}

Figure~\ref{fig:map} illustrates the area of interest for this case study, centered around Goslar. In \cite{bratzel2023flood}, the authors introduced a benchmark for predicting extreme hydrological events such as floods using a machine-learning algorithm, specifically a residual-LSTM model, within the Goslar region. Their model focused on predicting water levels at a specific sensor location, by leveraging data from multiple hydrological sensors. This study follows a similar benchmarking approach as outlined in \cite{bratzel2023flood}, ensuring comparability and methodological consistency. By aligning with the benchmark framework, we adopt a standardized evaluation protocol that facilitates direct performance assessment against existing state-of-the-art models. 

In addition to the primary study site, this paper examines \textbf{Göttingen}, a city located in southern Lower Saxony, Germany, as a secondary location. This selection is particularly relevant due to a significant flooding event that occurred in the Göttingen district on July 25\textsuperscript{th}, 2017. During this event, water levels rose to a peak of 140 cm \cite{flood_info_gott_neider_report}, causing widespread floods and damages. The primary hydrological source of the flood was the Innerste river, which originates in the Harz region and flows toward Hildesheim. This river is a 96.9 km long right tributary of the Leine in Lower Saxony, Germany, with a catchment area of 1,265.36 km\textsuperscript{2}. The flooding in Göttingen was a result of the failure of a 200-meter-long dam along the Innerste river, which led to extensive flooding throughout the affected area. To address the emergency, approximately 140 personnel were deployed to the flood-prone region to manage the crisis and support mitigation efforts. Their operations continued intensively until July 28, 2017, when the situation was brought under control, highlighting the severity and logistical challenges associated with the event. 

There are several challenges associated with physical sensors including high installation and maintenance costs, which result in limited spatial coverage, leaving many areas unmonitored. Even in regions where upstream data is present, the density may be insufficient, leading to data gaps. Additionally, technical malfunctions and physical damage can further result in missing or inaccurate data. In section \ref{model arc}, we will outline the architecture designed to address these limitations. The following section \ref{data} describes the datasets used in this paper.

\subsection{Data} \label{data}
This paper utilizes three datasets, one of which is the publicly available ground-based RADAR precipitation dataset from the DWD \cite{radar}, the national meteorological service of Germany. The dataset consists of Radar Online Calibration (RADOLAN) precipitation images \cite{dwd}, which are provided at a temporal resolution of 5-minute intervals and a spatial resolution of 1 km $\times$ 1 km. The selected time period for analysis spans from November 1, 2003, to June 30, 2018, covering approximately 15 years of continuous precipitation data. 

The second dataset comprises water level measurements for Goslar recorded at Sennhuette hydrological sensor station provided by Harzwasserwerke GmbH, corresponding to the same time interval. These water level measurements are captured at a temporal resolution of 15-minute intervals and serve as the target variable in our experiments for Goslar. 

The third dataset comprises water level measurements for Göttingen provided by Niedersächsischer Landesbetrieb für Wasserwirtschaft, Küsten- und Naturschutz (NLWKN) \cite{nlwkn}, corresponding to the same time interval with 15-minutes resolution. This dataset is used for validation purposes in this paper. The subsequent section details the steps employed for input and output preprocessing, which was applied identically to both locations, Goslar and Göttingen. While the described preprocessing steps specifically reference the target sensor Sennhuette (Goslar), these steps are equally applicable to the second target sensor in Göttingen.

\subsubsection{Input data preprocessing}
The precipitation data provided by the DWD covers the entirety of Germany, spanning approximately 1100 $\times$ 900 km. However, since our case study is focused on the region of Goslar, it is necessary to extract the relevant spatial subset from the larger precipitation images. This extraction is performed using a custom module that takes the geographical coordinates of Goslar as input, converts them into the corresponding indices within the precipitation image, and then clips a defined area based on the specified height and width. The output of this module is an index array, which is used to extract the relevant portion of the precipitation image. The resulting 2D array contains the rainfall values for the Goslar region, recorded at 5-minute intervals. 

\begin{figure}[H]
    \centering
    \includegraphics[width=0.5\textwidth]{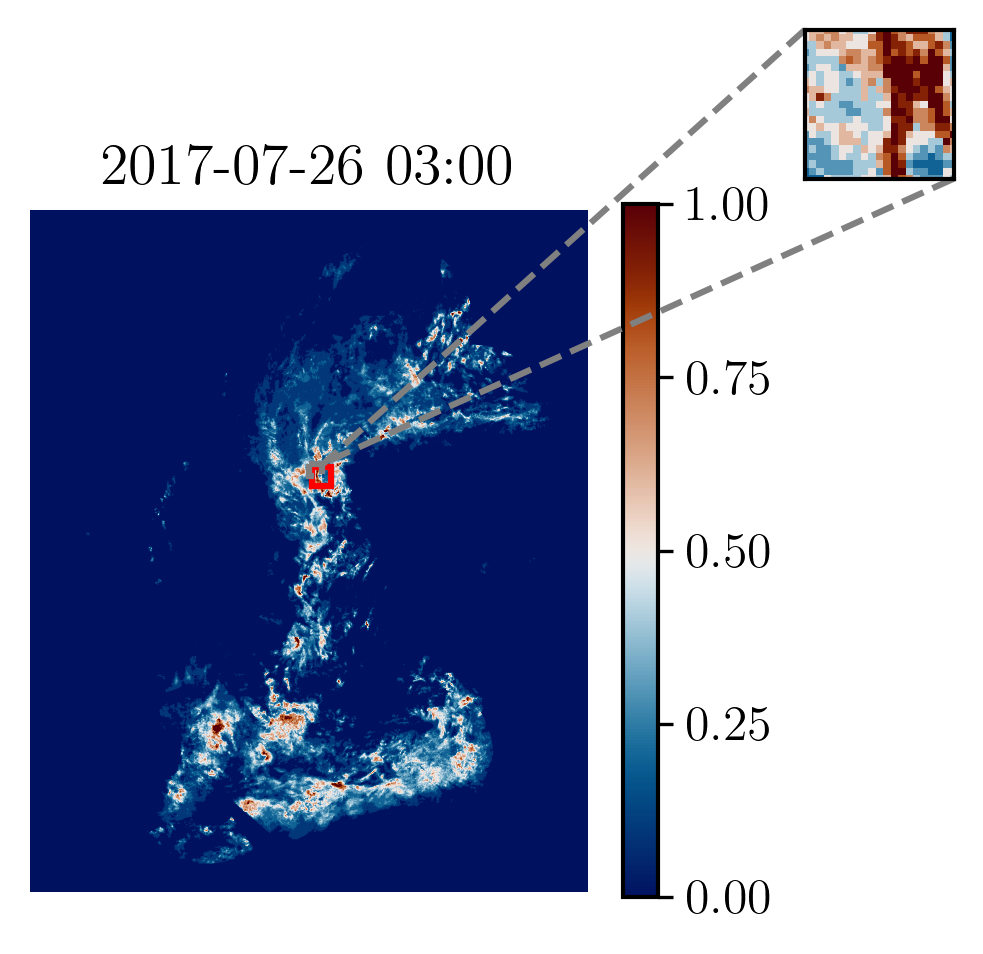}
    \includegraphics[width=\textwidth]{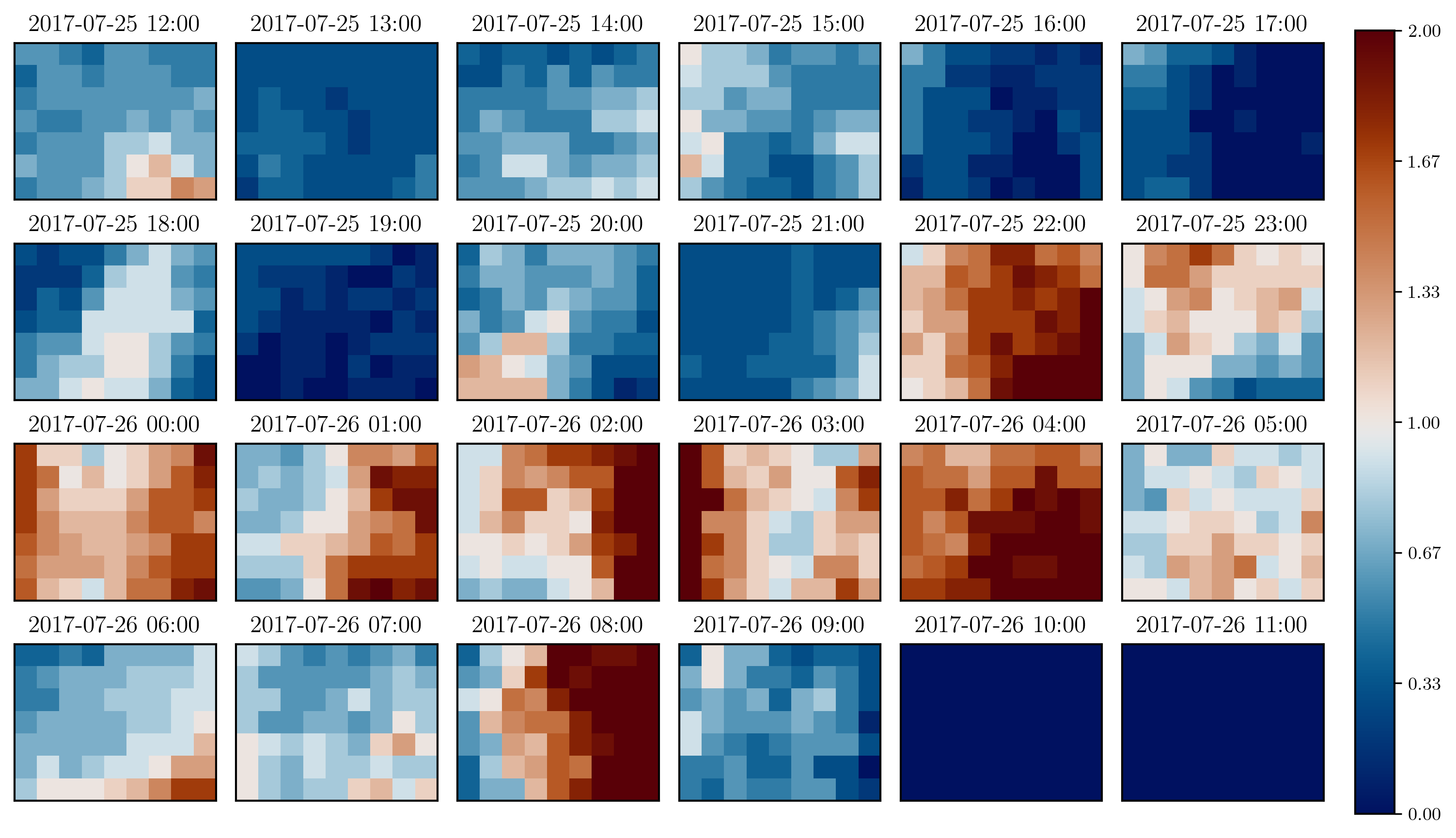}
    \caption{\textbf{Top}: A sample radar precipitation image over Germany from extreme flood event on 26\textsuperscript{th} July 2017 03:00 with a  resolution of 1km $\times$ 1km. The zoomed-in box shows the data for the Goslar region. \textbf{Below}: Hourly precipitation intensity maps for Goslar, blue indicating lower and brown indicating higher precipitation intensities.}
    \label{fig:radarsample}
\end{figure}

Figure~\ref{fig:radarsample} illustrates a sample precipitation image from flood event on 26$^{th}$ July 2017 in Goslar and hourly precipitation intensity maps for Goslar. To ensure temporal consistency with the target water level data, the precipitation values are aggregated by summing the data over three consecutive time steps, aligning with the 15-minute frequency of the water level measurements. This process is applied to all input precipitation images from November 2003 to June 2018.

\subsubsection{Output data preprocessing}
\begin{figure}[H]
    \centering
    \includegraphics[width = 1\linewidth]{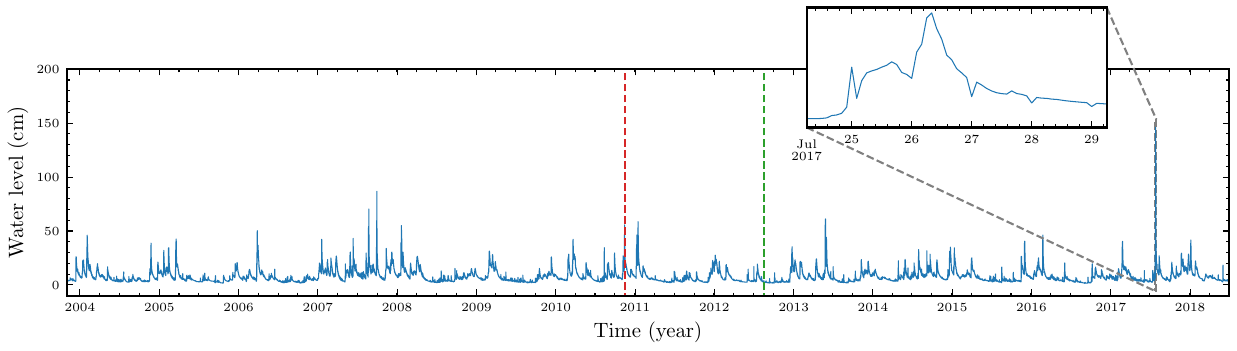}
    \caption{Water levels (cm) captured by a hydrological sensor at Sennhuette (until red line: training dataset, until green line: validation dataset and last part: testing dataset.}
    \label{fig:outputTS}
\end{figure}

The target water level time series, characterized by sparse flood events that cause sharp increases in water levels while the majority of values remain nearly constant, exhibits noise. To address this, the time series is smoothed using a simple moving average (SMA) with a window size of 8. Initially, the linear correlation between the cumulative precipitation data and the raw water levels at the Sennhuette station was found to be 0.075, indicating a weak relationship. However, after applying the smoothing technique, the correlation increased to 0.0835, demonstrating a slight improvement in the association between the input precipitation and water-level data. Figure~\ref{fig:outputTS} illustrates the output water level timeseries data used in the paper for Goslar location.

During preprocessing, it is observed that the relationship between precipitation and water levels exhibits non-linearity. Specifically, rainfall at time \textit{t} does not directly influence water levels at the same time point \textit{t}, but rather with a temporal lag. Analysis reveals that the change in water levels ($\Delta h$) over 2-hours intervals demonstrates a linear correlation coefficient of \textbf{0.335} with the cumulative precipitation data. This finding aligns with the expectation that the cumulative rainfall at a given location closely corresponds to the variation in water levels measured by a specific sensor. Consequently, the proposed model is designed to operate on the residuals of water levels, represented as $\Delta h$, rather than using the actual water levels ($h$) directly.

\begin{figure}[H]
    \centering
    \includegraphics[width = 1\linewidth]{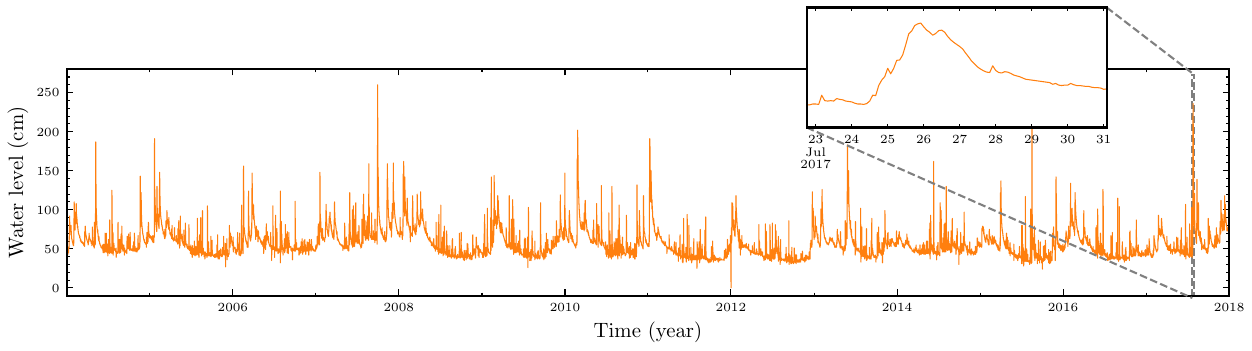}
    \caption{Water levels at Göttingen sensor station from 2003 to 2018.}
    \label{fig:TS_Göttingen}
\end{figure}

Figure~\ref{fig:TS_Göttingen} illustrates the output water level timeseries data used in the paper for Göttingen location. The input and output preprocessing was similarly conducted for Göttingen, revealing a linear correlation of 0.038 between the Göttingen sensor station data and the raw water levels. Upon re-evaluating the correlation using changes in water levels ($\Delta h_G$) over 2-hour intervals, an improved correlation coefficient of 0.255 was observed. This result clearly indicates that utilizing  $\Delta h_G$, is more appropriate than directly using the actual water levels ($h_G$) for Göttingen.

\section{Model Architecture} \label{model arc}
This section outlines the models used for experimentation and evaluation, by comparing four different models. The first is the residual LSTM (resLSTM) model, adapted from the flood prediction benchmark \cite{bratzel2023flood}. The second is a baseline model that assumes no change in water levels. The third is the STRPM model, a deep learning model that directly predicts water levels at the Sennhuette station ($h_s$). The fourth is the STRPM\textsubscript r model, a deep learning model that predicts the residuals of water levels ($\Delta h_s$) at the Sennhuette station.

The resLSTM model is a hybrid architecture combining a residual-LSTM network with an ANN, designed to forecast extreme flood events driven by sudden rainfall, as introduced in \cite{bratzel2023flood}. This model is trained to predict water levels at forecasting horizons of 2, 3, and 4 hours, using the preceding 96 timesteps (representing 24 hours of historical data). The resLSTM and the baseline models serve as benchmarks for assessing the performance of deep-learning models that utilize radar precipitation data.

The Baseline model provides a reference point for comparison by assuming no variation in water levels over the forecasting horizon. Its output is a simple projection of the current water level data, effectively shifting observed values forward in time without accounting for changes in the environment.

The STRPM and STRPM\textsubscript r models are both deep learning architectures that leverage radar-based precipitation data for water level prediction. The STRPM model directly predicts water levels ($h_s$) at Sennhuette, while the STRPM\textsubscript r model predicts the residuals of water levels ($\Delta h_s$). The residual-based approach in STRPM\textsubscript r is designed to focus on changes in water levels rather than absolute values, potentially improving performance in scenarios where small but critical fluctuations in water levels are essential for early flood detection.  

Let $\mathcal{N}_1$ and $\mathcal{N}_2$ represent STRPM and STRPM\textsubscript r models, $\boldsymbol{X}$ denote the set of input radar precipitation images, and $\boldsymbol{\Theta}_1$ and $\boldsymbol{\Theta}_2$ the learned model parameters. The estimated water level at time $t$ is expressed as:
\begin{equation}
    \hat{h} \textsubscript t = \mathcal{N}_1(\boldsymbol{X} ; \boldsymbol{\Theta}_1),
\end{equation}
where $\hat{h}_t$ represents the predicted water level. For STRPM\textsubscript r model, the estimated change in water level at time $t$ is expressed as:
\begin{subequations}
    \begin{equation}
        \Delta \hat{h}_ t = \mathcal{N}_2(\boldsymbol{X} ; \boldsymbol{\Theta}_2),
    \end{equation}
    \begin{equation}
    \hat{h}_\text{t} =  h_{\text{t}-\Delta \text{t}} + \Delta \hat{h}_t,
\end{equation}
\end{subequations}
where $\Delta \hat{h}_t$ is the predicted residual in water level between the two time steps and $h_{t-\Delta t}$ is the actual observed water level at a previous time step $t-\Delta t$. By predicting $ \Delta \hat{h}_t$ instead of $\hat{h}_t$ directly, the STRPM\textsubscript r is expected to better capture dynamic changes in water levels, potentially improving flood forecasting accuracy. The STRPM and STRPM\textsubscript r models employ a specialized (2+1)D convolution, which is detailed in the following section \ref{special convolution}, and the proposed model is described in section \ref{proposed model}.

\subsection{(2+1)D special convolution} \label{special convolution}
The Convolutional Neural Network (CNN) model utilized in this study is based on the approach outlined by Tran et al. \cite{tran2018closerlookspatiotemporalconvolutions}. The authors proposed an alternative approach that factorizes the 3D convolution into two sequential steps. The first step applies a convolution across the spatial dimensions only, with a shape of (1, height, width), followed by a second convolution along the temporal dimension, with a shape of (time, 1, 1). This decomposition significantly reduces the computational cost. 

\begin{figure} [h]
    \centering
    \includegraphics[width=1\linewidth]{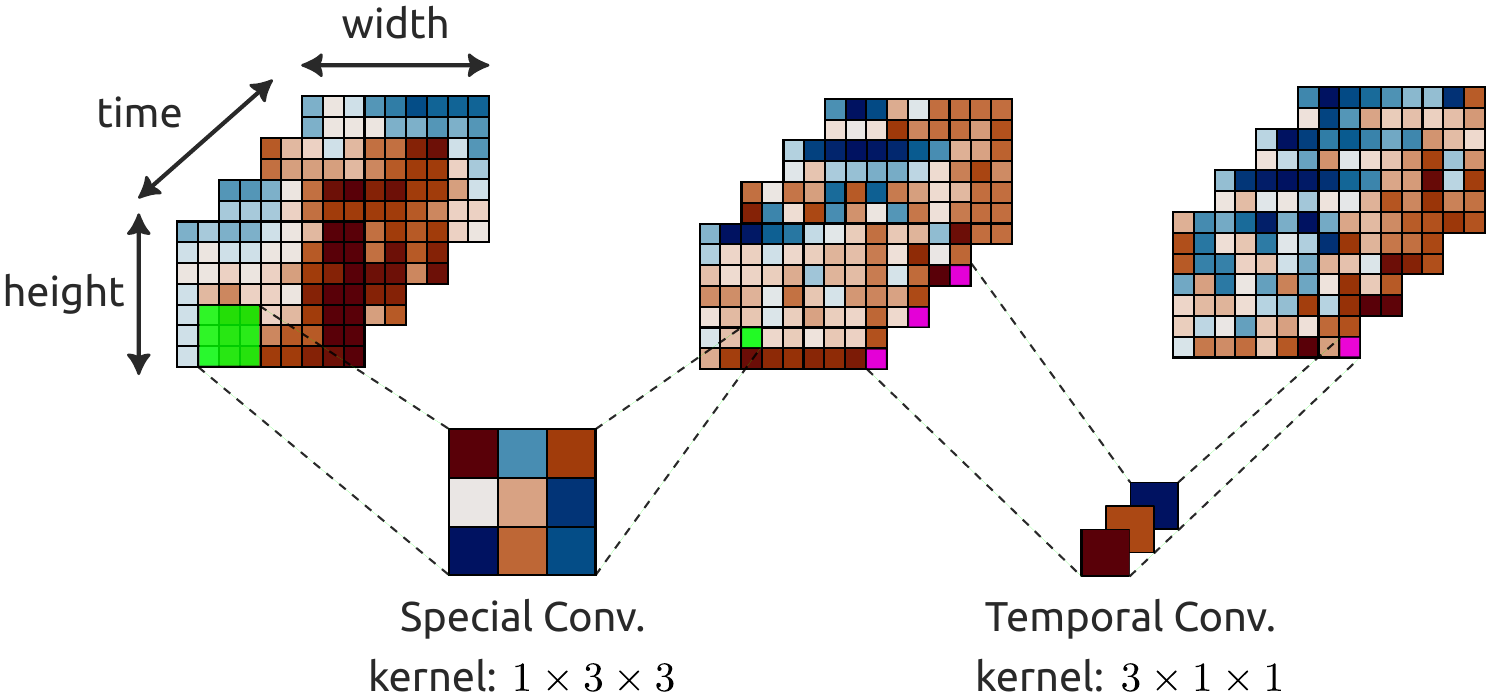}
    \caption{(2+1)D convolution architecture for a sample precipitation image from flood event on  26\textsuperscript{th} July 2017 in Goslar..}
    \label{fig:conv2plus1d}
\end{figure}

Figure~\ref{fig:conv2plus1d} provides a schematic illustration of the (2+1)D convolution operations. The figure contains a sample precipitation event from 26\textsuperscript{th} July 2017 at Goslar location. The parameter count for a (2+1)D convolution is (1 $\times$ height $\times$ width $\times$ channels) + (time $\times$ 1 $\times$ 1 $\times$ channels), effectively halving the number of parameters compared to a full 3D convolution (time $\times$ height $\times$ width $\times$ channels). For instance, a (2+1)D convolution with a kernel size of (3$\times$3$\times$3) requires (9$\times$channels\textsuperscript{2})+(3$\times$channels\textsuperscript{2}), which is significantly fewer parameters than those required by a full 3D convolution 27$\times$channels\textsuperscript{2}, reducing both the computational complexity and memory footprint of the model. In the next section \ref{proposed model}, we use this special convolution to build the proposed model architecture.
  
\subsection{Proposed model: STRPM\textsubscript r} \label{proposed model}
\begin{figure}[h]
\centering
\includegraphics[width = 1\linewidth]{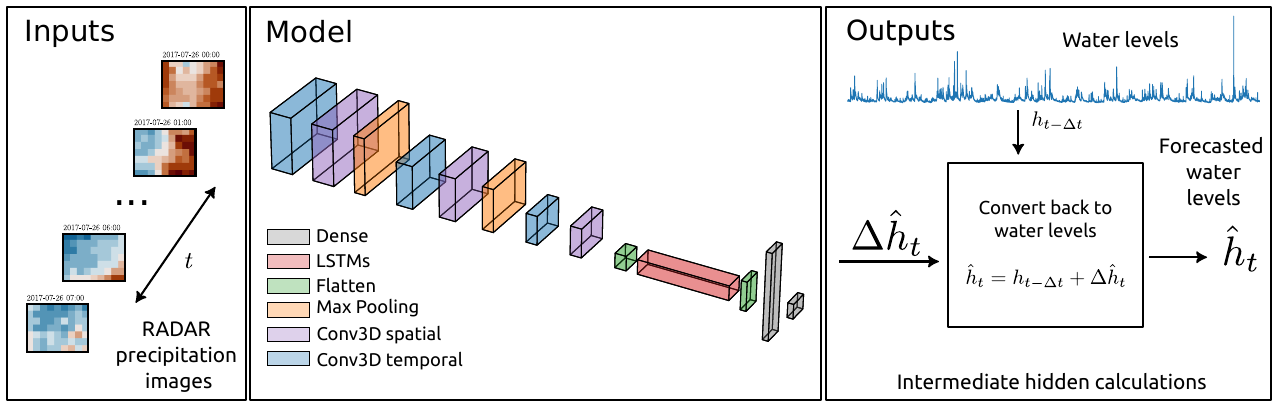}
\caption{Deep learning architecture for the Spatiotemporal Radar-based Precipitation Model for residual (STRPM\textsubscript r) model.}
\label{fig:model}
\end{figure}

The proposed model integrates a hybrid architecture that combines a specialized CNN with LSTM layers, specifically designed to process sequential radar precipitation data. The architecture is optimized for learning complex interactions between precipitation and flood response, and its structure is depicted in Figure~\ref{fig:model}.

In the proposed model, two-dimensional radar precipitation images are used as input and initially processed through a series of (2+1)D convolutional layers. After convolutional processing, the resulting feature maps are flattened, converting them into a sequential data format that can be effectively utilized by the LSTM layers. The LSTM layers model temporal dependencies within the data and predict future trends based on the processed precipitation input. One of the key features of this model is that the output of the LSTM layers corresponds to the residual water level changes ($\Delta \hat{h}_t$) at the Sennhuette sensor station. These residuals, representing time-based fluctuations in water levels, undergo further processing through additional network layers. In the final stage of the model, the residuals are transformed back into the original water levels ($\hat{h}_t$) via a reconstruction mechanism that restores the predicted water levels from their residual representation. This transformation process is illustrated within the intermediate hidden calculations block in Figure~\ref{fig:model}. The integration of special convolutions for efficient spatiotemporal feature extraction, combined with LSTM layers for temporal sequence modeling, ensures that this architecture is well-suited for flood prediction tasks that rely on radar-derived precipitation data. 

The architecture for the baseline, STRPM, and STRPM\textsubscript r models comprises four LSTM layers with hidden sizes of 128, 64, 32, and 8 units, respectively, followed by a fully connected layer. The models are trained over 50 epochs with a batch size of 256, using the Adam optimizer \cite{kingma2017adammethodstochasticoptimization}. A (2+1)D convolutional kernel of size 3$\times$3$\times$3 is applied to the input data. The training process utilizes mean squared error (MSE) as the loss function, while mean absolute error (MAE) serves as the evaluation metric. The activation function used for all intermediate layers is the hyperbolic tangent (tanh), while the final output layer employs a linear activation function.

The radar dataset with 514,176 precipitation images is split into 60\% for training and 40\% for testing, similar to the Benchmark paper \cite{bratzel2023flood}. The training set is further divided, with 80\% used for actual model training and 20\% for validation. Also in these experiments, each forecasting horizon incorporates past information from the previous 32 timesteps, equivalent to 8 hours to inform the predictions.  Notably, the extreme flooding event on 26\textsuperscript{th} July 2017, where water levels surged to 170.4 cm, is included in the test dataset to evaluate the model’s ability to predict rare and extreme events. This poses a challenge for generalizing predictions for this specific extreme event.

\begin{table}[h]
\small
\caption{Input and output data for all four models.}\label{tab:iodatainfo}
\resizebox{\textwidth}{!}{ %
\begin{tabular}{p{0.16\textwidth}p{0.32\textwidth}p{0.45\textwidth}}
\toprule
\textbf{Model}	& \textbf{Input}	& \textbf{Output}	 \\
\midrule
\textbf{resLSTM \cite{bratzel2023flood}} & Multiple sensors & Water levels at Sennheutte \\
\textbf{Baseline} & Current ground based sensor & Current ground based sensor as forecasts \\
\textbf{STRPM} & Precipitation images & Water levels at Sennheutte\\
\textbf{STRPM\textsubscript r} &Precipitation images & Residuals of water levels at Sennheutte \\
\bottomrule
\end{tabular}
}
\end{table}

The Table~\ref{tab:iodatainfo} illustrates the diverse range of inputs and outputs used by the four models in our study. The resLSTM \cite{bratzel2023flood} rely on traditional sensor data, while STRPM and STRPM\textsubscript r models leverage advanced radar-based precipitation images. By comparing these models, we aim to demonstrate the potential of radar-based approaches in enhancing water level prediction and flood forecasting accuracy. The models and experimental settings are described in the section \ref{results}.

\section{Results and Discussions} \label{results}
In this section, we present a detailed analysis of both the quantitative and qualitative results obtained from the experiments, followed by a comprehensive discussion of the key findings in each subsections. The quantitative results in section \ref{quantitative} are evaluated using a range of performance metrics, including MSE, Bravais-Pearson (BP), Nash-Sutcliffe Efficiency (NSE), and Index of Agreement (IoA) to assess the accuracy and robustness of the proposed models. The qualitative assessments in section \ref{qualitative} examine the model's performance on specific case studies, including extreme events, to evaluate the practical effectiveness of the predictions. Through visualizations of predicted versus actual water levels, we highlight the model's ability to capture rapid changes in water levels due to precipitation. Also, the model findings in terms of the broader implications for flood forecasting at another location Göttingen are described in section \ref{Göttingen}. 

\subsection{Quantitative results for evaluation framework} 
\label{quantitative}
In this section, the quantitative metrics for measuring the floods in Goslar are presented. These include MSE evaluations, overall metrics evaluations, and event focused evaluations, similar to \cite{bratzel2023flood}. 

\renewcommand{\arraystretch}{1.5}
\begin{table}[H]
\caption{Mean squared error (MSE) evaluations for different models for Goslar location.}
\label{tab:mse}
\small
\centering
\begin{tabular}[h]{p{0.18\textwidth}p{0.1\textwidth}p{0.1\textwidth}p{0.1\textwidth}p{0.1\textwidth}p{0.1\textwidth}}
\toprule
Model	& 2h & 3h &  4h & 8h & 12h\\
\midrule
STRPM & 39.2940 & 37.0361 & 40.6746 & 39.8148 & 39.4660\\
resLSTM \cite{bratzel2023flood} & 1.2007 & 1.208 & 1.841 & - &- \\
Baseline & 0.7306 & 1.1261 & 1.5230 & 3.2955 & 5.0322 \\
STRPM\textsubscript r & \textbf{0.7186} & \textbf{1.0592} & \textbf{1.0592} & \textbf{1.3548} & \textbf{2.8206} \\
\bottomrule
\end{tabular}
\end{table}

Table~\ref{tab:mse} compares the performance of four different models: STRPM, resLSTM \cite{bratzel2023flood}, Baseline, and STRPM\textsubscript r, across five time intervals: 2 hours, 3 hours, 4 hours, 8 hours, and 12 hours. The numerical values represent the performance metrics of each model at these intervals, i.e., the MSE. STRPM shows higher values across all time intervals compared to other models. This suggests that STRPM may have higher error rates or lower accuracy in this context. The STRPM\textsubscript r demonstrates lower error values compared to the baseline, highlighting its effectiveness in flood forecasting. The STRPM\textsubscript r appears to be the most accurate model, followed by the Baseline, with STRPM showing the lowest accuracy. This comparison is crucial for understanding the strengths and weaknesses of each model in predicting water levels and flood forecasting.

\renewcommand{\arraystretch}{1.5}
\begin{table}[H]
\centering
\caption{Overall metrics evaluation at different forecasting horizons for all traditional and deep learning models for Goslar region. STRPM: CNN-LSTM model for predicting water levels ($h_s$) directly, resLSTM model \cite{bratzel2023flood}, Baseline model, and STRPM\textsubscript r: CNN-LSTM model for predicting changes in water levels ($\Delta h_s$).}
\label{tab:metrics}
\small
\begin{tabular}{cc|ccc}
\toprule
Forecasts & Model/Metrics & BP & NSE & IoA \\
\midrule
\multirow{4}{*}{2h} 
  & STRPM & 0.170 & -0.0362 & 0.322\\ 
  & resLSTM \cite{bratzel2023flood} & 0.986 & 0.969 & 0.992 \\ 
  & Baseline & 0.990 & 0.980 & 0.995\\ 
  & STRPM\textsubscript r & \textbf{0.991} & \textbf{0.981} & \textbf{0.996} \\ \hline

\multirow{4}{*}{3h}   
  & STRPM & 0.217 & 0.023 & 0.319\\ 
  & resLSTM \cite{bratzel2023flood} & 0.985 & 0.969 & 0.991\\ 
  & Baseline &  0.985 & 0.970 &0.992 \\ 
  & STRPM\textsubscript r &  \textbf{0.986} & \textbf{0.972} & \textbf{0.993} \\ \hline

\multirow{4}{*}{4h}
  & STRPM & 0.152 & -0.072 & 0.332\\ 
  & resLSTM \cite{bratzel2023flood} & 0.979 & 0.953 & 0.991 \\ 
  & Baseline & 0.979 & 0.959 & 0.989\\ 
  & STRPM\textsubscript r &  \textbf{0.983} & \textbf{0.964} & \textbf{0.991}\\ \hline

\multirow{3}{*}{8h} 
  & STRPM & 0.134 & -0.049 & 0.278\\ 
  & Baseline & 0.956 & 0.913 & 0.977\\ 
  & STRPM\textsubscript r & \textbf{0.965} & \textbf{0.925} & \textbf{0.981}\\ \hline

\multirow{3}{*}{12h} 
  & STRPM & 0.191 & -0.041 & 0.332\\ 
  & Baseline & 0.933 & 0.867 & 0.965\\ 
  & STRPM\textsubscript r &  \textbf{0.951} & \textbf{0.898} & \textbf{0.975}\\ 
\bottomrule
\end{tabular}
\end{table}

The evaluation framework provided by authors in \cite{bratzel2023flood} is used in this section and the metrics are reported in Table~\ref{tab:metrics}. This table compares the performance of various models across different forecasting horizons (2 hours to 12 hours) and contains metrics such as BP correlation coefficient, NSE and IoA. BP measures the correlation between observed and predicted values. Across all forecasting horizons, model STRPM\textsubscript r consistently achieves the highest correlation, for instance, it yields a correlation of 0.991 at the 2h horizon and maintains strong performance as the forecasting horizon extends to 12h, with a correlation of 0.951. In comparison, STRPM shows significantly lower correlation values, particularly at longer horizons, dropping to 0.191 at 12h. NSE indicates how well the model predictions compare to the mean of observed data. Similar to the Bravais-Pearson correlation, the STRPM\textsubscript r demonstrates superior performance across all horizons. At 2h, the NSE for STRPM\textsubscript r is 0.981, maintaining strong predictive ability at 12h with an NSE of 0.898. The STRPM on the other hand, exhibits poor performance beyond 2h, with an NSE as low as -0.041 at the 12h horizons in terms of NSE. IoA assesses the degree of agreement between observed and predicted values, ranging from 0 (no agreement) to 1 (perfect agreement). The STRPM exhibits significantly lower IoA values, particularly as the prediction horizon increases, indicating less accurate predictions. The STRPM\textsubscript r outperforms the others, with an IoA of 0.996 at 2h and 0.975 at 12h.

\begin{table}[H]
\centering
\caption{Event focused evaluation framework for all four models (STRPM, resLSTM \cite{bratzel2023flood}, Baseline and STRPM\textsubscript r) at different forecasting horizons.}
\label{tab:evaluation_framework}
\small
\begin{tabular}{cc|ccccccc}
\toprule
Forecasts & Model/Metrics & T\textsubscript{ok} & T\textsubscript{over} & T\textsubscript{under} & error\textsubscript{average}  \\ 
\midrule
\multirow{4}{*}{2h}  
  & STRPM & 5 & 0 & 262 & 45.33 \\ 
  & resLSTM \cite{bratzel2023flood} & \textbf{253} & 18 & \textbf{26} & 27.60  \\
  & Baseline & 213 & 14 & 40 & 23.14 \\ 
  & STRPM\textsubscript r & 221 & \textbf{14} & 32 & \textbf{22.59} \\ \hline

\multirow{4}{*}{3h} 
  & STRPM &  8 & 0 & 259 & 43.81 \\ 
  & resLSTM \cite{bratzel2023flood}  & \textbf{233} & \textbf{14} & 50 & 24.20 \\ 
  & Baseline &  189 & 21 & 57 & 24.32 \\ 
  & STRPM\textsubscript r & 199 & 21 & \textbf{47} & \textbf{23.48} \\ \hline

\multirow{4}{*}{4h} 
  & STRPM & 3 & 0 & 264 & 45.21 \\
  & resLSTM \cite{bratzel2023flood}  & \textbf{222} & \textbf{15} & 60 & 26.3 \\ 
  & Baseline & 172 & 28 & 67 & 25.68 \\ 
  & STRPM\textsubscript r & 184 & 32 & \textbf{51} & \textbf{23.69} \\ \hline

\multirow{3}{*}{8h} & STRPM & 0 & 0 & 136 &  48.98\\ 
  & Baseline & 2 &0 & 265 & 46.32  \\ 
  & STRPM\textsubscript r & \textbf{136} & \textbf{38} & \textbf{93} & \textbf{26.51}\\ \hline

\multirow{3}{*}{12h} & STRPM & 6 & 0 & 130 & 40.57 \\ 
  & Baseline & 2 & 0 & 265 & 41.72 \\ 
  & STRPM\textsubscript r & \textbf{99} & \textbf{29}  & \textbf{139} & \textbf{27.19} \\
\bottomrule
\end{tabular}
\end{table}

In Table~\ref{tab:evaluation_framework}, the evaluation metrics focus on how well each model captures extreme events. The most important metrics are: T\textsubscript{ok}, the number of correct predictions where the predicted water levels matched the observed extreme event within an acceptable range (tolerance is $\pm$10 cm), T\textsubscript{over}, the number of instances where the model overestimated the water levels resulting in a potential false alert, T\textsubscript{under}, the number of instances where the model underestimated the water levels leading to a a potentially overseen right alert, and  error\textsubscript{average}, the absolute average deviation of the predictions from the observed values, reflecting the magnitude of prediction errors across events. For a more detailed event-focused metrics evaluated in this paper (for example, max\textsubscript{error}, T\textsubscript{not\textunderscore relevant}, annual\textunderscore events\textsubscript{over}, etc.), refer to \ref{appendix2}.

The results in Table~\ref{tab:evaluation_framework} demonstrate the comparative advantages of different models for forecasting water levels during extreme events. The STRPM\textsubscript r which predicts residual changes in water levels, consistently outperforms the STRPM model, especially at longer prediction horizons. The residual-based approach captures the dynamics of water level changes more accurately, as evidenced by fewer instances of underestimation and lower average errors across all horizons. The resLSTM model presented by Bratzel et al. \cite{bratzel2023flood}, is only up to 4 hours into the future, therefore Table~\ref{tab:metrics} and Table~\ref{tab:evaluation_framework} do not have those entries. The performance of resLSTM model for flood prediction seems to be comparative with the STRPM\textsubscript r in terms of T\textsubscript{ok} and T\textsubscript{under}. This implies that the proposed model can function as a virtual sensor by estimating water levels from radar precipitation data. This is useful for regions where physical sensors are unavailable or unreliable. Overall, the STRPM\textsubscript r model shows the most robust performance, particularly for medium to long-term flood predictions.
  
\subsection{Qualitative results for forecasting of extreme events}
\label{qualitative}
This section presents the forecasting results for the water level time series using the proposed STRPM\textsubscript r model. The model is capable of generating forecasts ranging from 2 hours to 12 hours into the future. 

\begin{figure}[H]
  \includegraphics[width=1\linewidth]{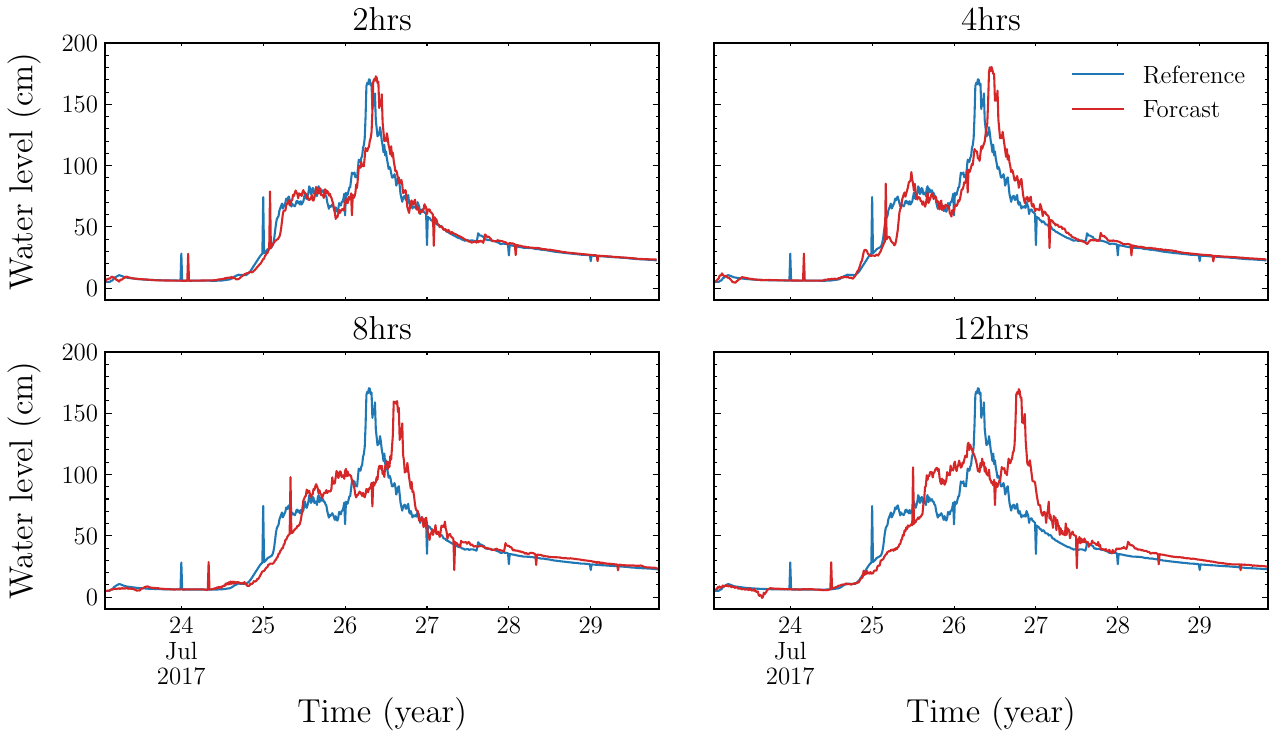}
  \caption{2 hours to 12 hours forecasts for the proposed model STRPM\textsubscript r, for an extreme flood event in Goslar from $23^{th}$ July, 2017 to $29^{th}$ July, 2017.}
  \label{fig:newforecasts}
\end{figure}

Figure~\ref{fig:newforecasts} presents a series of line graphs comparing observed and forecasted water levels over a seven-day period. Each panel corresponds to a different forecasting horizon, ranging from 2 hours to 12 hours. The y-axis represents water levels (in centimeters) recorded at Sennhuette, while the x-axis denotes the corresponding date and time. It illustrates that the forecasting model achieves reasonable accuracy for short-term predictions (2 to 4 hours), however, its accuracy diminishes for longer forecast horizons (8 to 12 hours), particularly during significant peak events. Despite this reduction in accuracy, the model demonstrates the capability to predict rising water levels 8 to 12 hours in advance, providing sufficient lead time for authorities to issue warnings and take appropriate measures. Since the extreme event is part of the testing dataset, extrapolation proved to be highly challenging, as the model had not encountered similar events during the training phase. Nevertheless, the proposed model still demonstrates promising predictive capabilities even for 12 hours, highlighting its potential for improving flood forecasting and early warning systems.

\begin{figure}[H]
  \includegraphics[width=\linewidth]{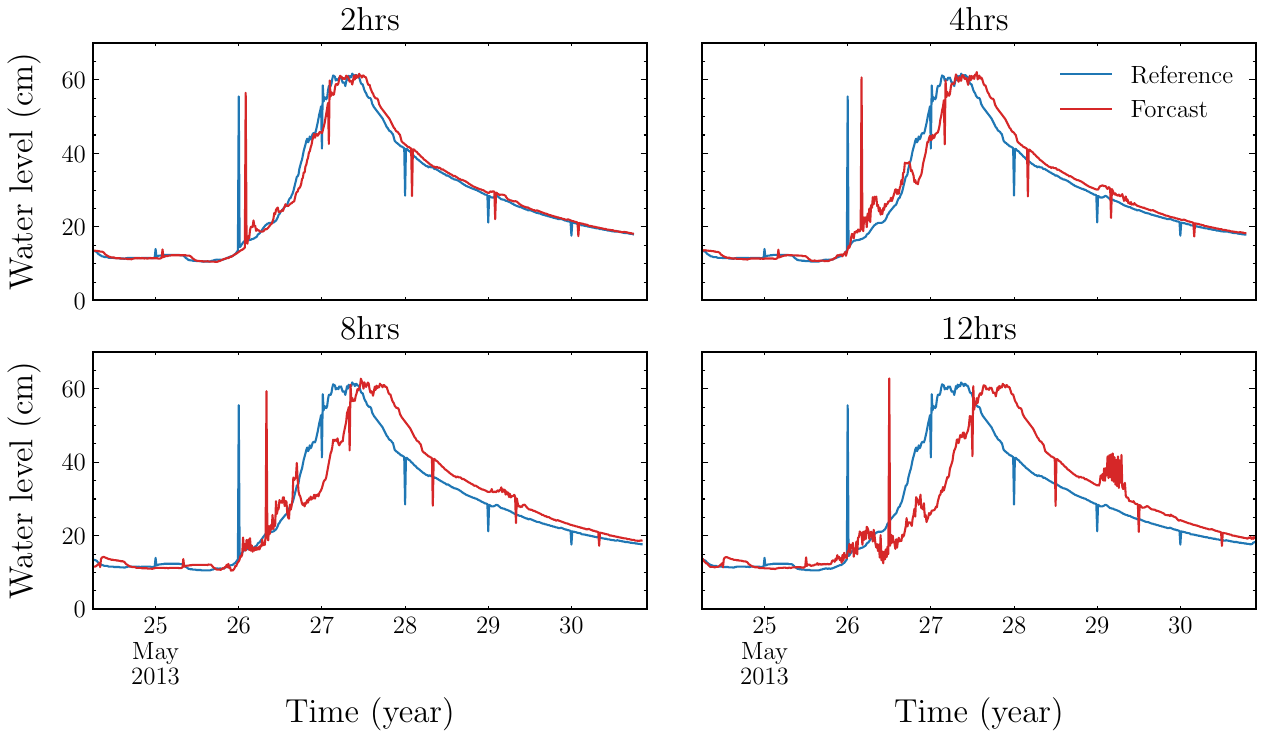}
  \caption{2 hours to 12 hours water level forecasts for the proposed model STRPM\textsubscript r, for timeperiod from $24^{th}$ May, 2013 to $30^{th}$ May, 2013.}
  \label{fig:forecasts}
\end{figure}

Figure~\ref{fig:forecasts} reflects water level forecasts for a different timeperiod from $24^{th}$ May, 2013 to $30^{th}$ May, 2013 for Goslar at Sennhuette. The short-term forecasting model (2 to 4 hours) demonstrates high accuracy and reliability, making it well-suited for immediate response and short-term planning. The long-term forecasting model (8 to 12 hours), while less reliable and possessing the lowest accuracy, is most effective for understanding general trends in water levels and preparing for potential future scenarios. 
\begin{figure}[h]
\centering
  \includegraphics[width=0.8\linewidth]{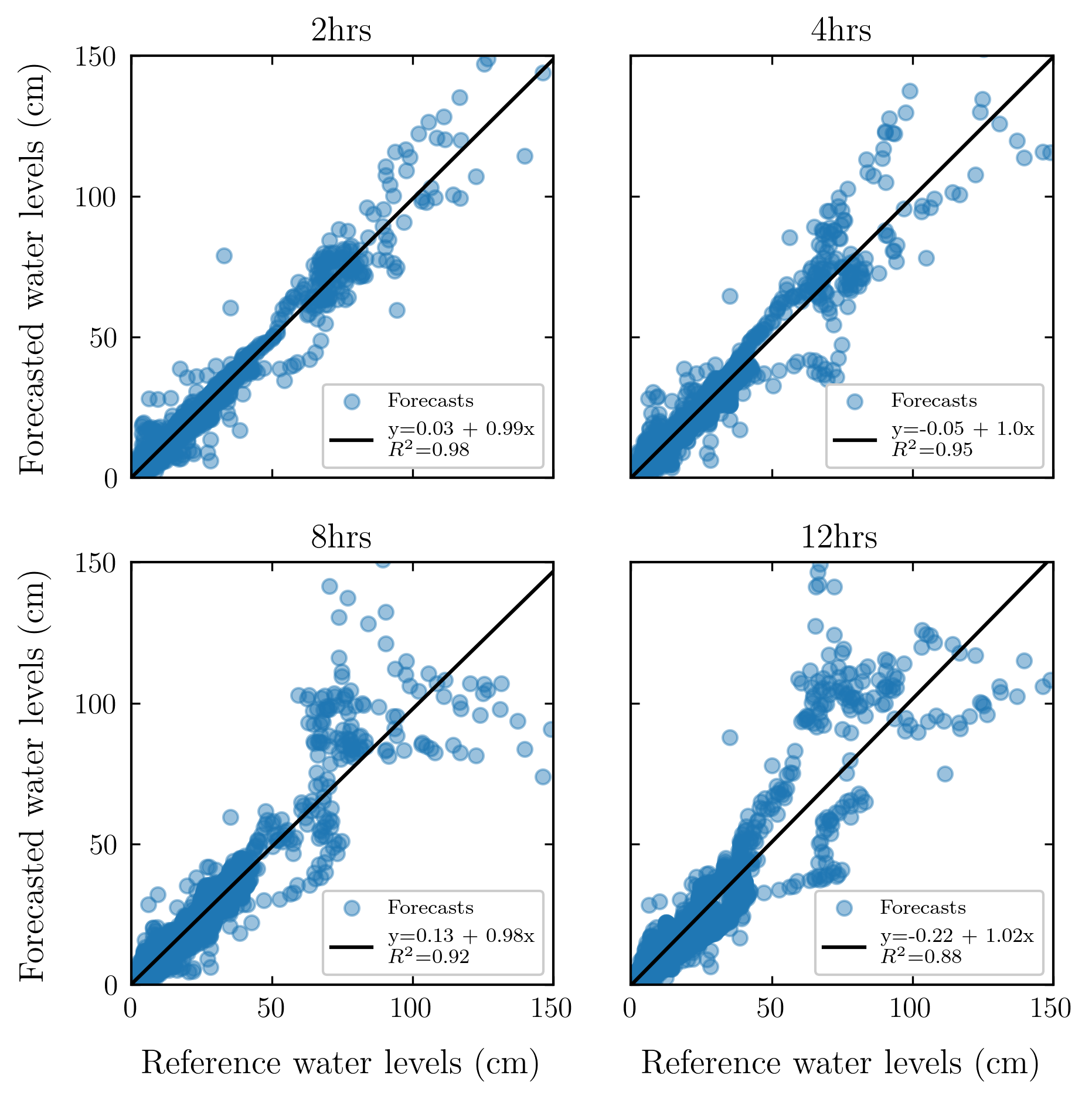}
  \caption{Scatter plot of water level forecasting at Sennhuette versus the actual water levels for STRPM\textsubscript r for Goslar.}
  \label{fig:scatterplot}
\end{figure}

Figure~\ref{fig:scatterplot} compares forecasted water levels against actual water levels (in cm) for various lead times for STRPM\textsubscript r model. Each plot includes a regression line with corresponding equations and $R^2$ values, which measure the fit of the forecasts to the observed data. The regression results indicate a high degree of accuracy, with $R^2$ values of 0.97616, 0.95092, 0.91758, and 0.88132 for 2-hour, 4-hour, 8-hour, and 12-hour forecasts, respectively. The model performs best for shorter lead times, as evidenced by the higher $R^2$ values and minimal deviations from the line of perfect agreement (y=x). However, the forecast accuracy decreases slightly as the lead time increases, showing a gradual reduction in $R^2$ and more noticeable deviations from actual water levels. This suggests that while the model reliably predicts short-term water levels, its performance diminishes with extended forecasting periods.

\subsection{Application to other regions}
\label{Göttingen}
\begin{figure}[h]
  \includegraphics[width=1\linewidth]{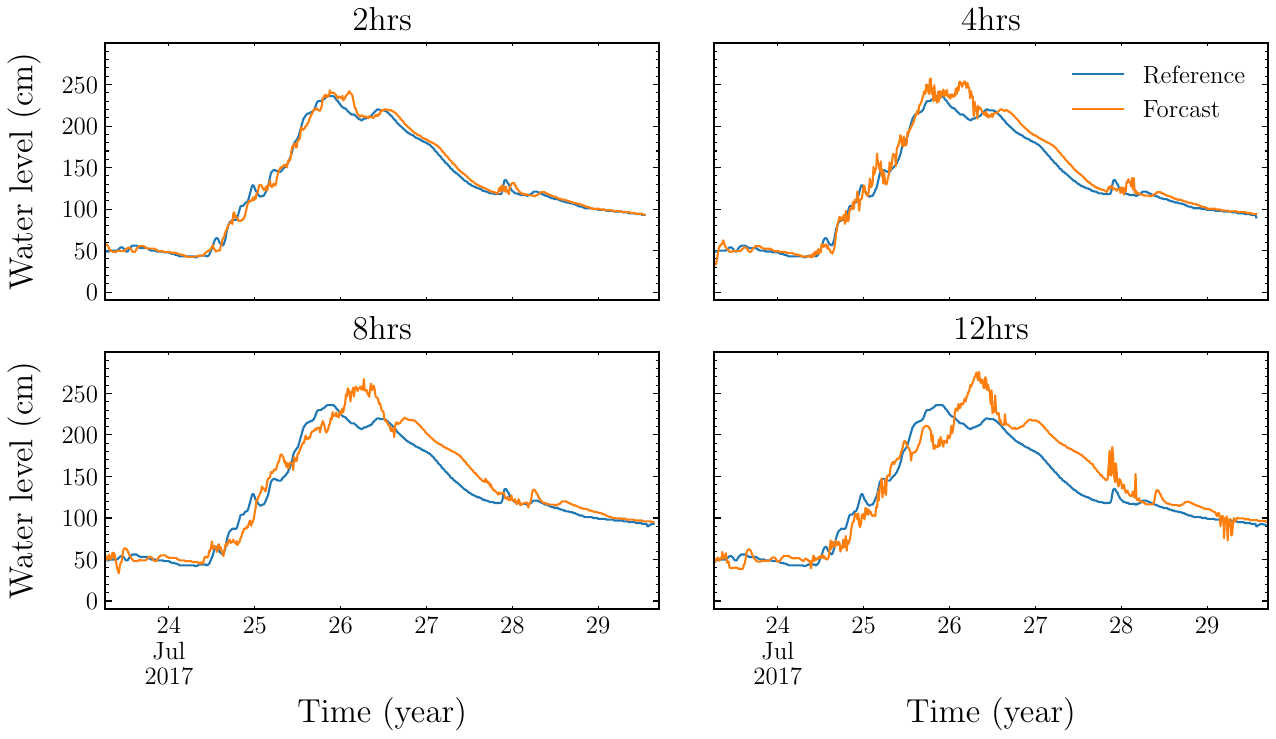}
  \caption{The reference water levels (blue) vs STRPM\textsubscript r (orange) water level forecasts for Göttingen for the extreme flood event on 25\textsuperscript{th} July, 2017 for 2- to 12-hours forecasting horizon.}
  \label{fig:Gottingen_forecasts}
\end{figure}

To validate the efficacy of the STRPM\textsubscript r for water level prediction, the model was re-trained using observational data specific to the Göttingen region. Ground-based sensor data for Göttingen provides accurate real-world measurements to serve as the ground truth for model evaluation. The validation outcomes, depicted in Figure~\ref{fig:Gottingen_forecasts}, offer a comparative analysis of actual observed water levels against the STRPM\textsubscript r forecasts across multiple forecasting horizons.

\begin{figure}[h]
  \includegraphics[width=1\linewidth]{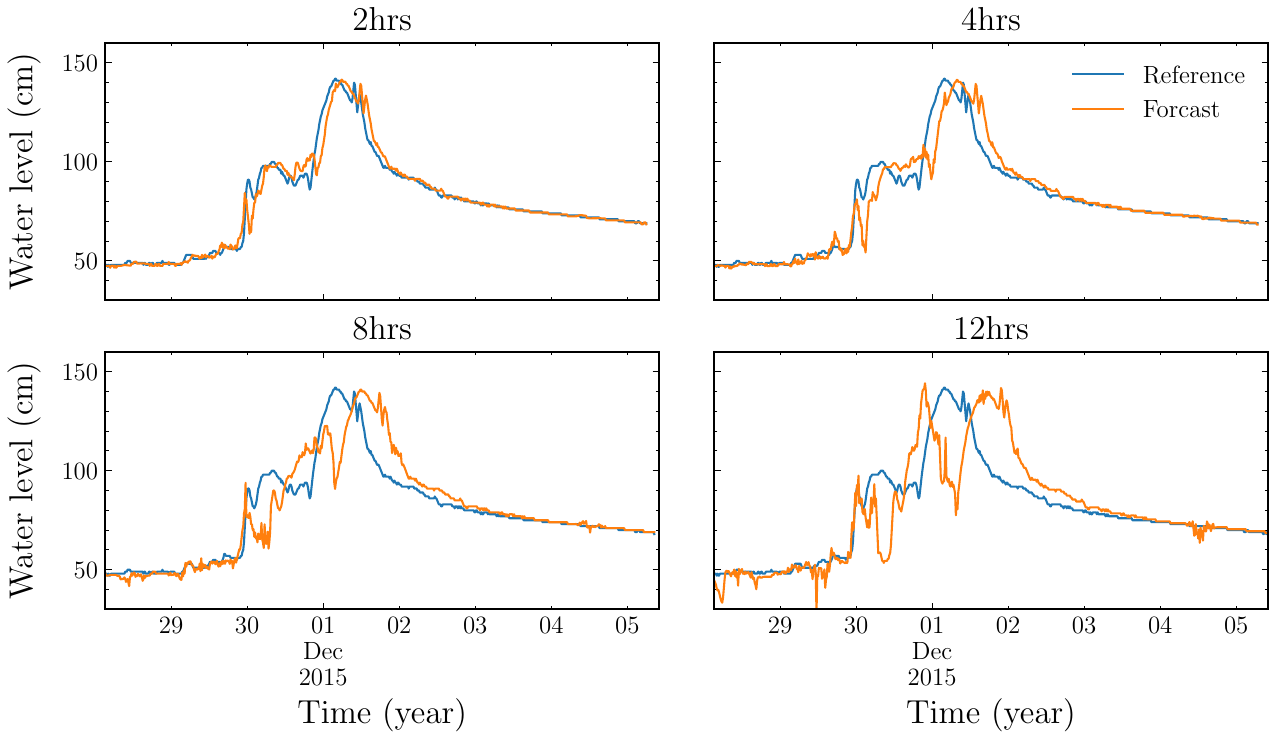}
  \caption{The reference water levels (blue) vs STRPM\textsubscript r (orange) water level forecasts for Göttingen for the second highest peak for 2- to 12-hours forecasting horizon.}
  \label{fig:Gottingen_secondpeak}
\end{figure}
Figure~\ref{fig:Gottingen_forecasts} depicts the forecasts for the highest peak of water levels in Göttingen and Figure~\ref{fig:Gottingen_secondpeak} depicts the second highest peak in Göttingen. In these figures, each subplot represents water level predictions over varying forecast horizons: 2 hours, 4 hours, 8 hours and 12 hours. The blue line signifies the actual recorded water levels from the NLWKN sensor data, serving as the ground truth reference. The accuracy of the proposed model STRPM\textsubscript r for Göttingen in the peaks of water levels is especially evident in the shorter forecast horizons (2–4 hours), where the model maintains high precision with the actual measurements. As the forecast horizon extends, STRPM\textsubscript r continues to show competitive performance, although with a slight increase in deviation from the actual values, but still able to capture the rise in the water levels.

\renewcommand{\arraystretch}{1.5}
\begin{table}[h]
\caption{Mean squared error (MSE) evaluations for Göttingen location.}
\label{tab:mse_Göttingen}
\small
\centering
\begin{tabular}[h]{p{0.15 \textwidth}p{0.1\textwidth}p{0.1\textwidth}p{0.1\textwidth}p{0.1\textwidth}p{0.1\textwidth}}
\toprule
Model & 2h & 3h	& 4h & 8h & 12h \\
\midrule
STRPM & 296.0928 & 305.5168 & 289.4087 & 287.981 & 290.3592\\
Baseline & 10.4279 & 14.2190 & 17.4329 & 30.4476 & 44.9132 \\
STRPM\textsubscript r & \textbf{9.818} & \textbf{12.1422} & \textbf{14.9846} & \textbf{24.6171} & \textbf{38.9187} \\
\bottomrule
\end{tabular}
\end{table}

The MSE for all three models for Göttingen is presented in Table~\ref{tab:mse_Göttingen}. It is evident that the STRPM\textsubscript r model achieves the lowest MSE across all forecasting horizons. This validation process not only confirmed STRPM\textsubscript r's predictive accuracy across varying forecast horizons but also provided insights into model refinement specific to regional hydrological patterns. The MSE for Göttingen in Table~\ref{tab:mse_Göttingen} is almost 10 times higher than that for Goslar in Table~\ref{tab:mse}. This is because of the hydrological precipitation patter difference between Göttingen and Goslar. The model performance is evaluated using raw water levels and both regions have different magnitudes of water level changes.  In the future, this model could be extended using a transfer learning approach, enabling the model to adapt its learned parameters from Göttingen to other regions with limited historical data. This would enhance STRPM\textsubscript r's applicability and robustness in diverse hydrological environments, facilitating scalable and region-specific flood forecasting solutions. In next section \ref{conclusion}, we conclude the result findings and recommend potential applications for future.


\section{Conclusions} \label{conclusion}
In this paper, we present a comprehensive regional study on forecasting water levels in Goslar and Göttingen, Germany, by only using radar-based precipitation data. The study is motivated by the extreme flood events in 2017 in both regions, where water levels rose to 1.59 meters in Goslar and 1.40 meters in Göttingen, causing severe flooding in their respective regions. This study involves analyzing spatiotemporal precipitation patterns derived from radar images and their relationship to hydrological sensors. The primary objective is to evaluate interactions between precipitation events and river water levels to enhance predictive flood modeling. The absence of upstream data makes this approach more generalizable, as it eliminates dependency on hydrological inputs that may not be available in all regions of Germany. This independence allows the proposed model to be applied in diverse hydrological settings, particularly in other regions of Germany with access to RADOLAN precipitation data.

One of the key contributions of this paper is the innovative approach to addressing the non-linear correlation between precipitation and water levels by predicting the residuals of water levels at sensors. By incorporating the residuals to the deep-learning models for Goslar, a significant improvement in linear correlation is observed, raising from 7\% to 33\% and 3\% to 25\% for Göttingen. The proposed model for predicting water levels using their residuals (STRPM\textsubscript r) demonstrates superior performance over the existing benchmark \cite{bratzel2023flood} and baseline model. This improvement is particularly pronounced for shorter-term forecasts, with the STRPM\textsubscript r showing strong predictive capabilities at 2-hour and 4-hour forecasting horizons, but also being able to predict 12 hours into the future by capturing the rise in the water levels. The performance metrics BP, NSE, and IoA for STRPM\textsubscript r for 2-hours horizon are 0.991, 0.981, and 0.996 and for 4-hours horizon are 0.983, 0.964, and 0.991, respectively. It can be concluded that the STRPM\textsubscript r outperforms the baseline and the event-focused evaluation framework further validates the robustness of the STRPM\textsubscript r in accurately forecasting extreme flood events, minimizing both over- and under-prediction errors. These results highlight the effectiveness and accuracy of the proposed approach.

The promising predictive capabilities of STRPM\textsubscript r are further validated by applying it to an additional location, Göttingen. By achieving consistent predictive capabilities at Göttingen without changing any model hyperparameters, the study demonstrates that STRPM\textsubscript r is not only extendable but also robust in its predictive accuracy. This highlights STRPM\textsubscript r's potential for broader applicability and adaptability to diverse datasets. Lastly, STRPM\textsubscript r can be used in the future for other tasks such as the imputation of missing observations within hydrological monitoring systems. By incorporating advanced techniques like transfer learning \cite{w16040607} and federated learning \cite{chen2023promptfederatedlearningweather}, the model can be extended to various network topologies, accommodating additional atmospheric parameters, such as temperature, humidity, and snowmelt. In conclusion, this study highlights the essential role of integrating precipitation data without the use of upstream data for extreme flood forecasting.

\section*{Author contribution}
All authors contributed to the study's conception and design. Material preparation and data collection were performed by \textbf{Sakshi Singh}. Project administration and supervision were carried out by \textbf{Stefan Wittek} and \textbf{Andreas Rausch}. Formal analysis and visualization were performed by Sakshi Singh and \textbf{Hamidreza Eivazi}. The first draft of the manuscript was written by Sakshi Singh and all authors commented on previous versions of the manuscript. All authors read and approved the final manuscript.

\section*{Funding}
We acknowledge support from the Open Access Publishing Fund of Clausthal University of Technology.

\section*{CRediT authorship contribution statement}
\textbf{Sakshi Singh}: Methodology, Visualization, Investigation, Writing – original draft, Writing – review \& editing. \textbf{Stefan Wittek}: Conceptualization, Validation, Writing – review \& editing, supervision. \textbf{Hamidreza Eivazi}: Validation, Writing – review \& editing. \textbf{Andreas Rausch}: Supervision, Funding acquisition.

\section*{Declaration of Competing Interest}
The authors declare that they have no known competing financial interests or personal relationships that could have appeared to influence the work reported in this paper.

\section*{Acknowledgments}
Our special thanks go to the Harzwasserwerke GmbH and NLWKN, for providing the data from the measuring stations in Goslar and Göttingen respectively. Special thanks to the city of Goslar for the great collaboration. 

\section*{Data Availability}
The codes and original sensor data presented in the study will be openly available upon acceptance. The RADOLAN precipitation data is freely available in the Open Data Server of the German Meteorological Service (DWD) at \cite{radar}.

\appendix
\section[\appendixname~\thesection]{Event focused framework results} \label{appendix2}
The event-focused framework discusses the danger events and allows the framework to be adjusted according to different hydrological conditions in Goslar. This section contains various metrics that indicate the rise in water level and help mitigate the situation of severe flooding. The minimum water level after which the measurements are relevant is 40 cm and the tolerance of the forecast (b) is 10 cm. For further details on these metrics, refer to paper \cite{bratzel2023flood}. The metrics are briefly described below: 
\begin{enumerate}[1.]
    \item T\textsubscript{relevant}: time points with rising or equal water level that is over the minimal level (40 cm)
    \item T\textsubscript{not\textunderscore relevant}: all events that are not relevant in T\textsubscript{relevant}
    \item T\textsubscript{ok}: time points where prediction is inside of tolerance (+/-b) relative to actual measurement
    \item T\textsubscript{under}: time points of underestimation of model (potential overseen right alert)
    \item T\textsubscript{over}: time points of overestimation of model (potential false alert)
    \item annual\textunderscore events\textsubscript{ok} = $\text{T}_\text{ok} / (\text{T}_\text{ok} + \text{T}_\text{under} + \text{T}_\text{over})$
    \item annual\textunderscore events\textsubscript{under} = $\text{T}_\text{under} / (\text{T}_\text{ok} + \text{T}_\text{under} + \text{T}_\text{over})$
    \item annual\textunderscore events\textsubscript{over} =  $\text{T}_\text{over} / (\text{T}_\text{ok} + \text{T}_\text{under} + \text{T}_\text{over})$
    \item annual\textunderscore events\textsubscript{all} = $(\text{T}_\text{ok} + \text{T}_\text{under} + \text{T}_\text{over}) / (\text{T}_\text{ok} + \text{T}_\text{under} + \text{T}_\text{over} + \text{T}_\text{not\textunderscore relevant})$
    \item T\textsubscript{ok\textunderscore average} [\%] = $\text{T}_\text{ok} / (\text{T}_\text{ok}+ \text{T}_\text{under}+ \text{T}_\text{over} + \text{T}_\text{not\textunderscore relevant})\times 100$
    \item T\textsubscript{under\textunderscore average} [\%] = $\text{T}_\text{under}/ (\text{T}_\text{ok}+ \text{T}_\text{under}+ \text{T}_\text{over} + \text{T}_\text{not\textunderscore relevant})\times 100$
    \item T\textsubscript{over\textunderscore relative\textunderscore average} [\%] = $\text{T}_\text{over} / (\text{T}_\text{ok}+ \text{T}_\text{under}+ \text{T}_\text{over} + \text{T}_\text{not\textunderscore relevant})\times 100$
    \item error\textsubscript{sum}: absolute sum of errors = $abs|\text{error}_\text{under}|$ + $abs|\text{error}_\text{over}|$
    \item error\textsubscript{average}: absolute average of errors 
    \item error\textsubscript{max}: max of errors
    \item error\textsubscript{median}: median of errors
\end{enumerate}

The detailed event focused framework results for all four model for Goslar are presented in Table~\ref{tab:event_based_table_full}.

\begin{sidewaystable}
\renewcommand{\arraystretch}{1.7} 
\caption{Event focused evaluation framework for all four models in Goslar location. STRPM: CNN-LSTM model for predicting water levels ($h_s$) using precipitation directly, resLSTM \cite{bratzel2023flood}, Baseline model, and STRPM\textsubscript r: CNN-LSTM model for predicting changes in water levels ($\Delta h_s$) using precipitation.}
\label{tab:event_based_table_full}
\footnotesize
\centering
\tabcolsep=0.05cm
\begin{tabular}{c|cccc|cccc|cccc|ccc|ccc}
\toprule
\textbf{Metrics/Models} & \multicolumn{4}{|c|}{\textbf{2h}} & \multicolumn{4}{|c|}{\textbf{3h}} & \multicolumn{4}{|c}{\textbf{4h}} & \multicolumn{3}{|c|}{\textbf{8h}} & \multicolumn{3}{|c}{\textbf{12h}} \\ 
\midrule
 & \textbf{a} & \textbf{b} & \textbf{c} & \textbf{d} & \textbf{a} & \textbf{b} & \textbf{c} & \textbf{d} & \textbf{a} & \textbf{b} & \textbf{c} & \textbf{d} & \textbf{a} & \textbf{c} & \textbf{d} & \textbf{a} & \textbf{c} & \textbf{d} \\ \hline
 
T\textsubscript{not\textunderscore relevant} & 205368 & 205334 & 205360 & 205360 & 205366 & 205330 & 205354 & 205354 & 205364 &  205326 & 205348 & 205348 & 205358 & 205326 & 205326 & 205352 & 205304 & 205304 \\ \hline
T\textsubscript{ok}  & 5 & 253 & 213 & 221 & 8 & 233 & 189 & 199 & 3 & 222 & 172 & 184 & 2 & 116 & 136 & 2 & 82 & 99\\ \hline
T\textsubscript{over} & 0 & 18 & 14 & 14 & 0 & 14 & 21 & 21 & 0 & 15 & 28 & 32 & 0 & 30 & 38 & 0 & 25 & 29 \\ \hline
T\textsubscript{under}  & 262 & 26 & 40 & 32 & 259 & 50 & 57 & 47 & 264 & 60 & 67 & 51 & 265 & 121 & 93 & 265 & 160 & 139 \\ \hline
T\textsubscript{ok\textunderscore average} [\%] & 0.002 & 0.123 & 0.103 & 0.107 & 0.003 & 0.113 & 0.091 & 0.096 & 0.001 & 0.108 & 0.083 & 0.089 & 0.0009 & 0.056 & 0.0564 & 0.0009 &  0.039 & 0.048 \\ \hline
T\textsubscript{under\textunderscore average} [\%]  & 0.127 &  0.0126 & 0.019 & 0.0155 & 0.125 & 0.0243 & 0.0277 & 0.0228 & 0.128 & 0.0292 & 0.0325 & 0.0248 & 0.128 & 0.0588 & 0.0452 & 0.128 & 0.077 & 0.067\\ \hline
T\textsubscript{over\textunderscore relative\textunderscore average} [\%] & 0 & 0.0088 & 0.006 & 0.0068 & 0 & 0.0068 & 0.0102 & 0.0102 & 0 & 0.0073 & 0.0136 & 0.0155 & 0 & 0.0145 & 0.018 & 0 & 0.012 & 0.0141  \\ \hline
annual\textunderscore events\textsubscript{all}  & 45.496 & 50.6 & 45.498 & 45.498 & 45.496 & 50.6 & 45.499 & 45.499 & 45.497 & 50.6 & 45.500 & 45.501 & 45.498 & 45.505 & 45.505 & 45.499 & 45.510 & 45.510\\ \hline
annual\textunderscore events\textsubscript{ok}   & 0.851 & 43.1 & 36.29 & 37.65 &1.363 & 39.7 & 32.207 & 33.911 & 0.511 & 37.8 & 29.311 &  31.356 & 0.340 & 19.77 &23.17 & 0.340 & 13.97 & 16.87 \\ \hline
annual\textunderscore events\textsubscript{over}   & 0 & 3.1 & 2.385 & 2.385 & 0 & 2.4 & 3.578 & 3.578 & 0 & 2.6 & 4.771 & 5.453 & 0 & 5.11 & 6.476 & 0 & 4.261 & 4.943 \\ \hline
annual\textunderscore events\textsubscript{under}  & 44.644 & 4.4 & 6.816 & 5.452 & 44.133 & 8.5 & 9.713 & 8.00 & 44.985 & 10.2 & 11.417 & 8.691 & 45.157 & 20.622 & 15.85 & 45.159 & 27.272 & 23.69 \\ \hline
error\textsubscript{sum} & 11877.39 & 1214.6 & 1249.6 & 1039.21 & 11348.56 & 1549.2 & 1897.5 & 1596.58 & 11936.26 & 1972.5 & 2440 & 1966.6 & 12274.13 & 4623.1 & 3473.55 & 11056.52 & 6140.5 & 4568.83\\ \hline
error\textsubscript{average} & 45.333 & 27.6 & 23.14 & 22.591 & 43.816 & 24.2 & 24.326 & 23.479 & 45.213 & 26.3 & 25.684 & 23.694 & 46.317 & 30.616 & 26.515 & 41.722 & 33.191 & 27.195 \\ \hline
error\textsubscript{max} & 138.949  & 84.3 & 64.4 & 60.85 & 134.494 & 77.5 & 77.3 &  73.05 & 133.303 & 84.6 &  79.9 & 67.89 & 153.157 & 100.1 & 84.67 & 128.913 & 102.5& 99.37\\ \hline
error\textsubscript{median} & 37.221 &  21.9 & 18.65 & 18.04 & 35.608 & 19.2 & 20.4 & 18.14 & 37.114 & 21 & 19.9 & 18.30 & 38.280 & 20.7 & 19.07 & 33.608 & 25.1 & 19.79\\ 
\bottomrule
\end{tabular} 
\end{sidewaystable} 

\begin{table*}[!ht]
 \begin{framed}
\renewcommand{\nomname}{Nomenclature} 
\input{nomenclature.nls}
\end{framed}
\end{table*}


\end{document}